\documentclass{webofc}
\usepackage[varg]{txfonts} 
\usepackage[capitalise]{cleveref}
\def\aap{A\&A}%
\def\apj{ApJ}%
\def\jcap{JCAP}%
\def\mnras{MNRAS}%
\def\pasa{PASA}%
\begin{document}
\title{Multipole analysis of cluster weak lensing shear in {\sc The Three Hundred} project}

\author{\lastname{C. Payerne}\inst{1}\fnsep\thanks{e-mail: constantin.payerne@lpsc.in2p3.fr} \and 
\lastname{C. Hanser}\inst{\ref{LPSC}} \and
       \lastname{C. Murray}\inst{\ref{LPSC},\ref{APC}} \and \lastname{N. Amouroux}\inst{\ref{LPSC},\ref{Lapp}}  \and
         \lastname{C. Combet}\inst{\ref{LPSC}}  
}
\institute{Université Grenoble Alpes, CNRS-IN2P3, LPSC, 38000 Grenoble, France \label{LPSC} \and 
Université Savoie Mont Blanc, CNRS-IN2P3, LAPP, 74941 Annecy-le-Vieux, France\label{Lapp}
\and Université Paris Cité, CNRS-IN2P3, APC, 75013 Paris, France \label{APC}}

\abstract{Weak gravitational lensing is an important tool to estimate the masses of galaxy clusters, as it allows us to directly access their projected surface mass density along the line-of-sight (LOS) in a manner largely independent of their dynamical state. Moreover, we can extract information on the projected shape of the cluster mass distribution. In this work, we generate mock catalogs of lensed background galaxies to measure the individual lensing properties of galaxy clusters from the simulation project \textsc{The Three Hundred}. By repeating the analysis for different projections of the same cluster, we find that the use of shear multipoles provides constraints on the ellipticity of the cluster projected mass density but does not have a significant impact on the cluster mass reconstruction compared to the standard approach.}
\maketitle
\section{Introduction}
\label{intro}
Galaxy clusters are not expected to be spherical, due to the non-trivial shape of the initial density peaks from which they originate, and due to their complex individual accretion history in the cosmic web. They are shown to have a complex morphology in simulations \cite{Despali2014triaxialsim}, with a preference for being prolate rather than oblate\footnote{Most of them look like rugby balls rather than an athletics discus.} but also in real data, for instance through the distribution of cluster member galaxies \cite{BBinggeli1982triaxialclustergalaxies}, the X-rays surface brightness \cite{Lau2012triaxialXrays} or weak lensing. For instance, Oguri et al. (2010) \cite{Oguri2010ellipticity} measured the average axis ratio\footnote{$q=b/a$, where $b$ and $a$ are the minor and major axis of the elliptical projected mass density.} $q = 0.54\pm 0.04$ for 25 X-rays selected clusters analyzing the HSC two-dimensional shear maps.

In the context of cluster count cosmology, the weak lensing mass is generally estimated assuming spherical symmetry (see e.g. \cite{Bocquet2019SPT}). However, this hypothesis may lead to a significant bias in the lensing mass reconstruction. Simulation-based weak lensing studies (see e.g. \cite{Herbonnet2022the300bias,munozecheverría2023galaxy}) show that the lensing mass estimates are significantly impacted by halos' elliptical shapes, specifically by the combination of elongation and orientation of the mass distribution that may lead to a bias of $30\%$ to the recovered mass.

Cluster triaxiality is therefore a source of systematic uncertainty in the interpretation of the weak lensing signal, thus on which may in turn impact cluster-based cosmological analyses. Being able to detect asphericity and measure robust masses on individual galaxy clusters is a prerequisite for taking full advantage of the next generation of cluster cosmology data.

When the projected dark matter distribution is not spherical, the cluster's shear field can be analyzed by studying its several multipole moments. Clampitt et al. in \cite{Clampitt2016lensingellipticity} and Robison et al. in \cite{Robison2023multipole} have measured the ellipticity of the galaxy-sized dark matter halos around SDSS Luminous Red Galaxies by analyzing the stacked lensing shear multipoles. From a cluster perspective,  Gouin et al. in \cite{Gouin2017multipolesaperture} have measured the multipolar moments of weak lensing signal around clusters in the dark-matter-only N-body PLUS
simulations to quantify the topology of the cosmic web. Gonzalez et al. in \cite{Gonzalez2020lensingellipticity} have measured the average projected ellipticity of SDSS redMaPPer galaxy clusters from stacked shear multipoles and found $\langle \varepsilon\rangle = (1-q)/(1+q)  = 0.21 \pm 0.04$ and Shin et al. in \cite{Shin2018stackedmultipolessdss} analyzed jointly the shear multipolar moments and the distribution of member galaxies to find $\langle \varepsilon \rangle =0.28\pm 0.07$. 

In this work we aim at inferring individual galaxy cluster parameters (such as the mass or the projected cluster ellipticity) by analyzing the shear multipole moments, in the context of upcoming wide galaxy surveys such as the Rubin Legacy Survey of Space and Time (LSST) or the \textit{Euclid} survey. We present in \cref{sec:multipole_formalism} the shear multipole formalism in cluster fields, and the corresponding lensing analyses of simulated galaxy clusters from the project {\sc The Three Hundred} in \cref{sec:app_300}. Then, we conclude in \cref{sec:conclusion}.

\section{Multipolar decomposition of the cluster shear field}
\label{sec:multipole_formalism}
The observed ellipticity $\epsilon^{\rm obs}$ of a background galaxy with intrinsic shape $\epsilon^{\rm int}$ is related to the reduced shear $g$ by 
\begin{equation}
    \epsilon^{\rm obs} =
        \frac{\epsilon^{\rm int} + g}{1 + g^*\epsilon^{\rm int}},
        \label{eq:lensing_eq}
\end{equation}
where the reduced shear $g = \gamma / (1 - \kappa)$ ($g^*$ is the complex conjugate), $\gamma$ and $\kappa$ are respectively the shear and the convergence. We consider the tangential and cross ellipticity $\epsilon_+ = -\mathrm{Re}[\epsilon \mathrm{e}^{-2i\varphi}]$ and $\epsilon_\times = -\mathrm{Im}[\epsilon \mathrm{e}^{-2i\varphi}]$ where $\varphi$ is the polar angle relative to the cluster center. The estimator of the $m$-th multipole moment (real part\footnote{The imaginary part is obtained by replacing the cosine with the sine function.}) of the excess surface density (ESD) is given by \cite{Gonzalez2020lensingellipticity} 
\begin{align}
\label{eq:deltasigma_multipole_Re}
     \widehat{\Delta\Sigma}_{\rm +/\times, \Re}^{(m)} &= \frac{1}{\sum\limits_{s = 1} w_{ls}(\cos m\varphi_s)^2}
     \sum\limits_{s= 1}w_{ls}\Sigma_{{\rm crit}}^{s,l}\epsilon_{+/\times}^{ls}\cos(-m\varphi_s),
\end{align}
where $\Sigma_{{\rm crit}}^{s,l}$ is the critical surface mass density between the cluster and the source galaxy, and $w_{ls}$ are individual lens-source weights. The ESD multipoles can reveal a wealth of information about the halo morphology \cite{Bernstein2009multipole}. Each $m$-th ESD multipole moment is sensitive to $\Sigma^{(m)}$, where $\Sigma(R, \varphi)$ is defined as the cluster projected surface mass density. Assuming spherical symmetry, we find that $\Sigma^{(m)} = \Delta\Sigma^{(m)}_{+/\times} = 0$ if $m\neq 0$, and only the monopole coefficients $\Delta\Sigma_+^{(0)}$ and $\Sigma^{(0)}$ remain. However, when $\Sigma$ is not just a pure radial function, several multipoles are not null.

Considering that the halo is well-approximated by a triaxial ellipsoid \cite{Knebe2006triaxialmodeling}, the corresponding projected surface density is given by \cite{Adhikari2015multipole} 

\begin{equation}
    \Sigma(R, \varphi) = \Sigma_{\rm sph}\left(R\sqrt{q\cos^2(\varphi-\varphi_0) + \frac{\sin^2(\varphi-\varphi_0)}{q}}\right).
    \label{eq:sigma_ell_partial}
\end{equation}
where $\Sigma_{\rm sph}$ is a spherical projected mass density profile. With $\epsilon = (1 - q^2)/(1 + q^2)$, we find that $\Sigma^{(2)}$ (by extension $\Delta\Sigma^{(2)}$) is proportional to $\epsilon\cos(2\varphi_0)$.

\section{Shear multipole analysis of {\sc The Three Hundred} galaxy clusters}
\label{sec:app_300}
\subsection{{\sc The Three Hundred} lensing dataset}

\textsc{The Three Hundred} (\textsc{The300}) project \cite{Cui2018The300} provides a set of 324 simulated galaxy clusters with masses $M_{\rm 200m} \in [6.4, 26.5]\times10^{14} M_\odot$. These clusters are the results of full-physics hydrodynamics simulations of the densest regions in the Multi-Dark Planck 2 N-body simulation \cite{Klypin2016MDsim}. The shear and convergence maps for each cluster have been produced\footnote{The shear maps have been derived from the equation of the deflection angle using the \textsc{The300} convergence maps. The procedure is detailed by Herbonnet et al. in \cite{Herbonnet2022the300bias} (see their Section 2.1.2) and is based on \cite{Meneghetti2010Xrays, Meneghetti2020lensing}.} for three different orthogonal projections along the line of sight\footnote{These orthogonal projections are \textit{randomly} oriented with respect to specific halo axes.}.  
In this work, we consider a random subset\footnote{The analysis will be further expanded to the full dataset.} of 40 clusters at redshift $z_l = 0.33$.

\subsection{Creating LSST-like background galaxy catalogs}
We generate mock galaxy catalogs, representative of what will be provided by the next-generation galaxy surveys, such as the Rubin Legacy Survey of Space and Time (LSST). First, we generate a sample of unlensed galaxies with a homogeneous galaxy number density of 30 gal\;arcmin$^{-2}$ and a redshift distribution from Chang et al. in \cite{Chang2013pdfgalaxies}. Intrinsic ellipticities are generated with a shape noise of $\sigma = 0.25$.

\begin{figure*}
\centering
\includegraphics[width = 1\textwidth]{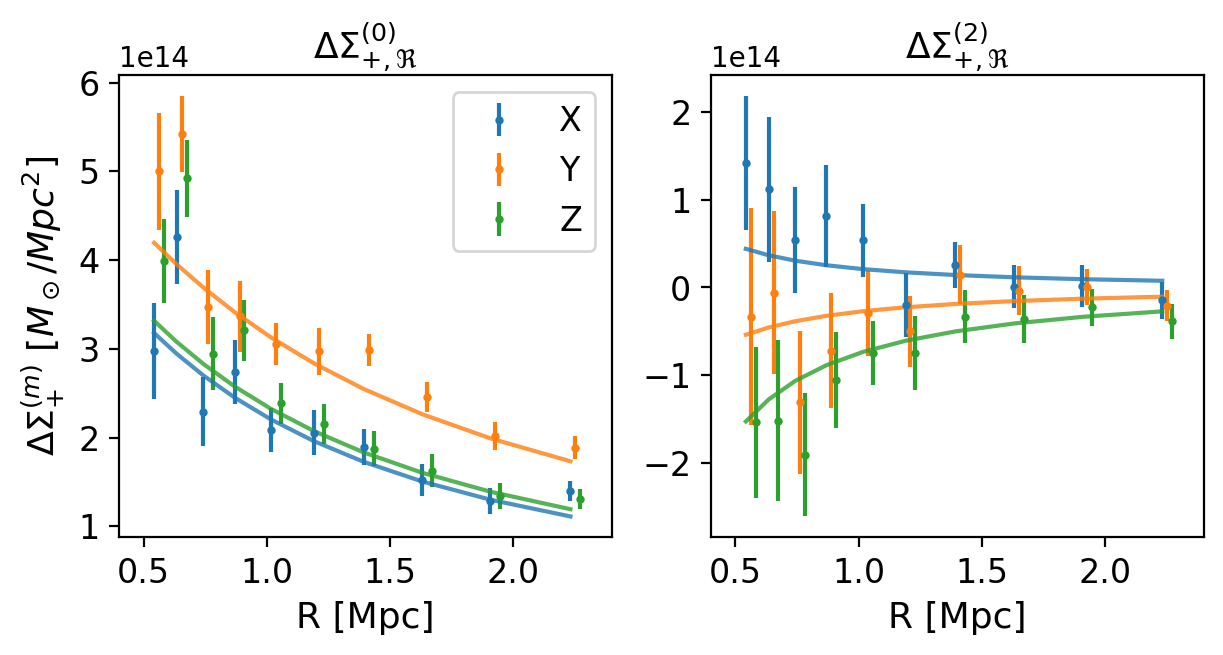}
\caption{Estimated profiles (points and error bars) and best fits (full lines) for $\Delta\Sigma_{+,\Re}^{(0)}$ (monopole, left panel) and for $\Delta\Sigma_{+,\Re}^{(2)}$ (quadrupole, right panel) according to the three different orthogonal projections along the LOS (called X, Y and Z). See text for details. }
\label{fig:DS_data}
\end{figure*}

Second, for each source galaxy, we use \cref{eq:lensing_eq} to derive the sheared ellipticity from its intrinsic shape by interpolating the \textsc{The300} lensing shear $\gamma_{\rm The300}(z_0)$ and convergence $\kappa_{\rm The300}(z_0)$ maps at the galaxy's position. Since the lensing maps are obtained at source redshift $z_0 = 3$, we compute the lensing maps for different source redshifts according to the "single-lens" approximation, such that\footnote{Again, we follow the methodology in \cite{Herbonnet2022the300bias} (see their Section 2.2).} the lensing maps $\gamma(z_i)$ and $\kappa(z_i)$ at source redshift $z_i$ are re-scaled from their $z_0$ values by the factor $\Sigma_{{\rm crit}}(z_l, z_0)/\Sigma_{{\rm crit}}(z_l, z_i)$, where $z_l$ is the cluster redshift. For each cluster, we have generated one lensed source catalog per orthogonal projection.

\subsection{Shear multipole analysis on a single cluster}

For a given cluster in our \textsc{The300} sample, we estimate the ESD multipole moments $m=0,2$ (monopole+quadrupole) from the mock source sample using \cref{eq:deltasigma_multipole_Re} in 10 radial bins from 0.5 to $\sim 2.5$~Mpc (maximum size of the aperture around the cluster centers). The corresponding data vectors for the three orthogonal projections are represented in blue ($X$ projection), orange ($Y$), and green ($Z$) in \cref{fig:DS_data}. The first result is that with LSST-like statistics (level of shape-noise, number density, and source redshift distribution), it is possible to measure the quadrupole signals for individual massive clusters, although with a lower signal-to-noise ratio than the monopole. The $\Delta\Sigma_{+,\Re}^{(0)}$ lensing profile for the $Y$ projection is higher in amplitude compared to those obtained with the two other projections. Moreover, the corresponding $\Delta\Sigma_{+,\Re}^{(2)}$ (in orange, right panel) is smaller in amplitude compared to the two others. The multipole moments for the two other orientations (in blue and green, right panel) show significant positive and negative values revealing their respective orientation. As the three orientations describe the same cluster seen from different angles, then we can reasonably state that for a prolate cluster, the $Y$ case corresponds roughly to the semi-major axis aligned along the LOS (the standard lensing signal is boosted, with no traces of projected ellipticity) and the two others cases to the semi-major axis perpendicular to the LOS (lowered monopoles, and respectively positive/negative quadrupole).

\begin{figure}
\centering
\includegraphics[width = 0.49\textwidth]{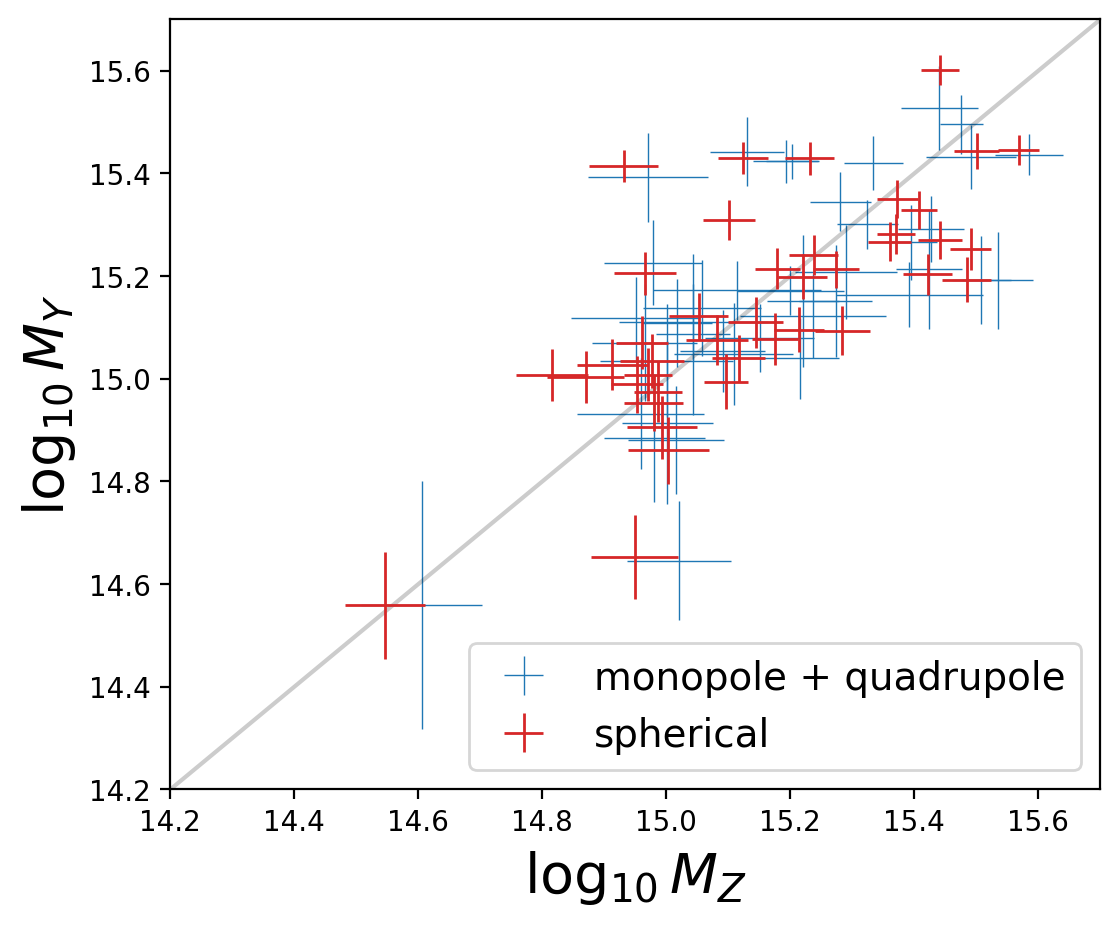}
\includegraphics[width = 0.5\textwidth]{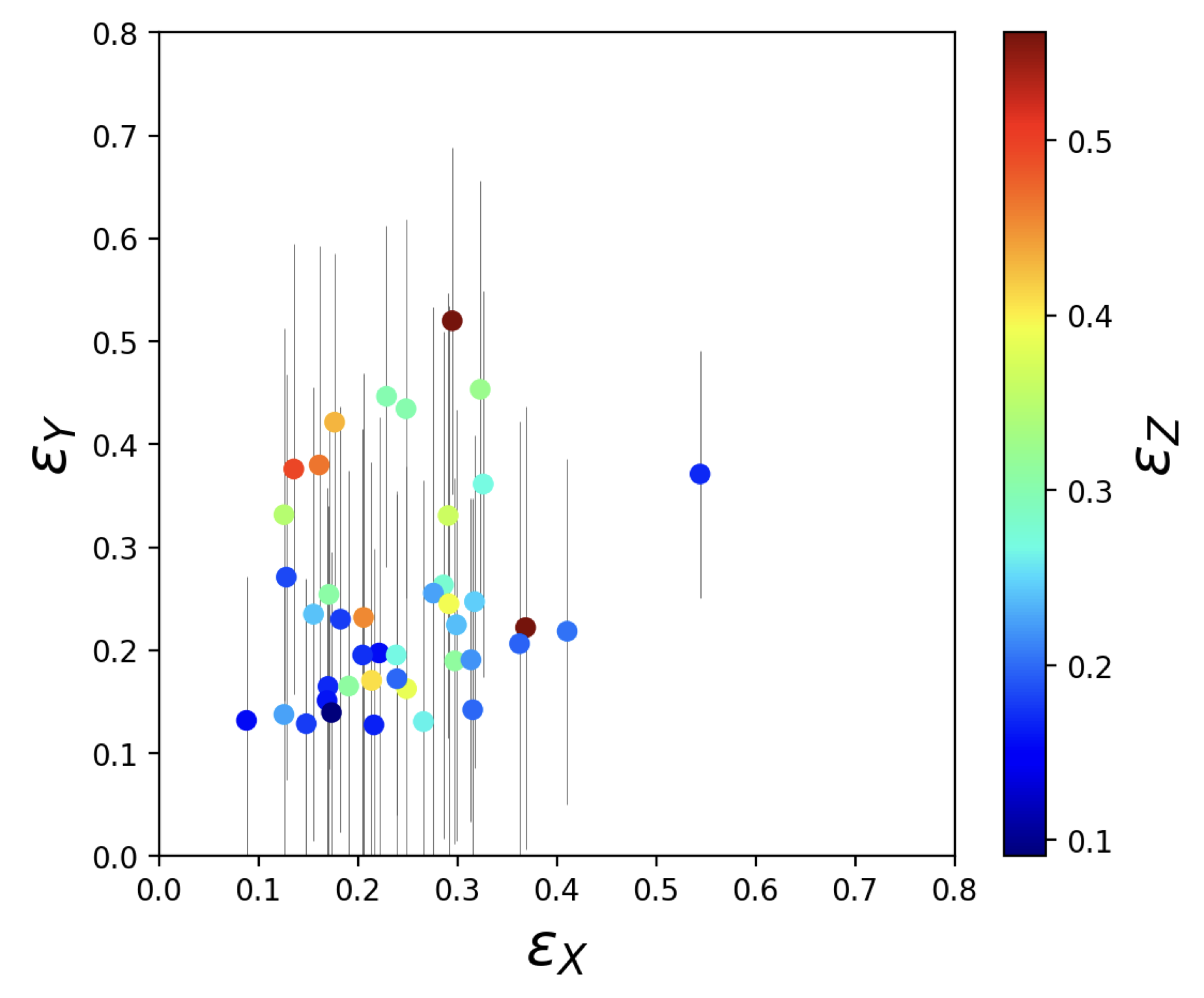}
\caption{{\it Left}: Lensing mass estimates for the projection $Y$ versus that obtained for the projection $Z$. Blue points: using monopole+quadrupole and elliptical modeling. Red points: using monopole and spherical modeling. {\it Right}: Ellipticity estimates for the $Y$ projection versus that obtained for the $X$ projection. The color bar on the right indicates the ellipticity estimates on the $Z$ projection. We only show the error bars of $\epsilon_Y$ for readability.} 
\label{fig:mass}       % Give a unique label
\end{figure}

To fit the cluster mass and ellipticity, we use a spherical Navarro-Frank-White dark matter profile \cite{Navarro1997nfw}, a concentration-mass relation from  \cite{Bhattacharya2013cM}, and then derive the elliptical projected density with \cref{eq:sigma_ell_partial}. To draw the parameter posterior we use a Gaussian likelihood for the observed profiles, where we estimate the joint covariance matrix using jackknife re-sampling \cite{Escoffier2016jackknife}. We do not account for the not-so-weak lensing correction in the modeling \cite{Mandelbaum2006wlcorrection} occurring at small radii, but that may be important for massive halos and may lead to an overestimation of the lensing mass.
We use a flat prior for the (log)mass between 14 and 16. We restrict the axis ratio to $q \leq 1$ (fixing $a$ to be the major axis). In this preliminary study, we only consider the real part of $m=2$ multipole of the tangential ESD\footnote{We could also use the imaginary part, that has other dependency with respect to the orientation angle $\varphi_0$. In this approach, $\varphi_0$ is a free parameter, opposite to the works in \cite{Gonzalez2020lensingellipticity,vanUitert2017lensingmultipole} that estimated the average multipoles around several lenses by stacking on a preferred orientation (the elongated axis of the projected mass density probed by the orientation of the BCG or the spatial distribution of the member galaxies). This rotation allowed them to consider $\varphi_0 = 0$, then using that $\Sigma^{(m)}_{\Im} = \Delta\Sigma_{\rm +, \Im}^{(m)} = 0$.}, thus being invariant under the transformation $\varphi_0' \rightarrow \varphi_0 + \pi$ and $\varphi_0' \rightarrow  \pi-\varphi_0$ so we use a flat prior within $[0, \pi/2]$. 

We show the corresponding best fit profiles in full lines in \cref{fig:DS_data}. We find the masses $\log_{10}M_X = 15.48\pm 0.04$, $\log_{10}M_Y = 15.77\pm 0.03$ and $\log_{10}M_Z = 15.53\pm 0.03$, and for the axis ratios we find the values $q_X = 0.66 \pm 0.11$, $q_Y = 0.80\pm 0.08$ and $q_Z = 0.51 \pm 0.07$. We find that the tension between the $M_Y$ and other masses is higher than 2$\sigma$. We see that the cluster seems prolate-shaped with major axis aligned along the $Y$ axis, since the monopole signal is boosted and there is no significant quadrupole information (the axis ratio $q_Y$ is closer to 1). For the two other projections, $\Sigma$ is more elliptical. We see that in this analysis setup, it is difficult to disentangle between a massive spherical cluster and a less massive ellipsoidal cluster with semi-major axis aligned along the LOS.

\subsection{Results on 40 galaxy clusters}

We repeat this analysis with 40 clusters. In \cref{fig:mass} (left panel), we compare the recovered $M_Y$ and $M_Z$ masses obtained with the approach above (blue points, monopole + quadrupole and elliptical modeling) to the ones obtained from the \textit{standard} approach (red points, monopole and spherical modeling\footnote{The modeling of the excess surface density profile for a spherical halo is given by $\Delta\Sigma_{+,\Re} = \langle \Sigma(<R)\rangle - \Sigma(R)$, where $\Sigma$ is the radial projected mass density profile.}).

If the mass reconstruction did not suffer from projection effects, every point should lie on the $x=y$ line (in black), i.e., the mass would be the same independent of the projection ($M_X = M_Y = M_Z$). However, due to the complex shape of the dark matter halo, each projected lensing mass is different. We note that using shear multipole + elliptical modeling (blue), the compatibility between per-projection lensing masses is improved since mass posteriors are more overlapping compared to the monopole + spherical modeling case (red); this is, however, due to larger error bars resulting from the added degrees of freedom (ellipticity and orientation angle) rather than an improvement in the mass. So, we find that the mass is not better constrained when using the multipole analysis, but the error bars are more realistic since they take into account the systematics due to deviation from the spherical hypothesis.

Finally, we define the average projected ellipticity per projection defined by $\varepsilon = (1-q)/(1+q)$. The ellipticity estimates per projection are shown in \cref{fig:mass} (right panel). We find that from the multipole analysis, the average ellipticities per projection are $\langle \varepsilon_{\rm X}\rangle = 0.24 \pm 0.08$, $\langle \varepsilon_{\rm Y}\rangle = 0.25 \pm 0.10$, and $\langle \varepsilon_{\rm Z}\rangle = 0.28 \pm 0.11$ compared to only fitting the monopole and assuming an elliptical modeling\footnote{For an elliptical halo, the tangential ESD monopole is given by $ \Delta\Sigma^{(0)}_{+,\Re} = \langle\Sigma^{(0)}(<R)\rangle - \Sigma^{(0)}(R)$, where $\Sigma^{(0)}$ is the averaged projected mass density profile.} that provided, for instance, $\langle \varepsilon_{\rm X} \rangle = 0.32 \pm 0.07$. The value inferred with multipoles is more compatible with other estimates of the mean cluster ellipticity, such as $\langle \varepsilon \rangle =0.28\pm 0.07$ found in \cite{Shin2018stackedmultipolessdss} from a stacked approach. These values still need to be compared to the ellipticity measured on the simulated projected maps.

\section{Summary and conclusions}
\label{sec:conclusion}
Galaxy clusters can be studied through the complex feature of their weak gravitational lensing shear field. In this work, we evaluate the benefit of using a multipole analysis of the lensing signal to infer the mass and ellipticity of individual clusters. This is done using the three orthogonal projections available in the cluster dataset from the simulation project \textsc{The Three Hundred} and generating LSST-like mock galaxy catalogs. We have shown that with the next generation of weak lensing surveys, we can obtain robust ellipticity measurements for relatively low-mass clusters\footnote{With current weak lensing catalogs, we can obtain robust individual cluster mass measurements for hundreds of galaxy clusters (see e.g., \cite{Murray2022}) and individual ellipticity measurements for the most massive clusters \cite{Oguri2010ellipticity}.} through measurement of the shear quadrupole, and at larger scales than that probed by other baryonic tracers, such as the orientation of the Brightest Central Galaxy. However, using this information, we did not find a strong impact on the recovered weak lensing mass compared to the standard approach, although it better accounts for systematics related to deviation from sphericity in the recovered errors.

\end{document}